\documentclass{ws-ijmpa}

\usepackage[super]{cite}
\usepackage{color, amsmath, amssymb, graphicx, textcomp, amsfonts, graphics, epstopdf, slashed, multirow, adjustbox, lipsum, rotating, xcolor, wrapfig, epsfig, ulem, tikzsymbols, tikz}
\usepackage[verbose,hypertexnames=false]{hyperref}
\hypersetup{colorlinks=false,allbordercolors=blue,pdfborderstyle={/S/U/W 1}}

\definecolor{lime}{HTML}{A6CE39}
\DeclareRobustCommand{\orcidicon}{
	\begin{tikzpicture}
	\draw[lime, fill=lime] (0,0)
	circle [radius=0.2]
	node[white] {{\fontfamily{qag}\selectfont \tiny ID}};
	\draw[white, fill=white] (-0.0625,0.095)
	circle [radius=0.007];
	\end{tikzpicture}
	\hspace{-2mm} }
\foreach \x in {A, ..., Z}{\expandafter\xdef\csname orcid\x\endcsname{\noexpand\href{https://orcid.org/\csname orcidauthor\x\endcsname}
			{\noexpand\orcidicon}} }

\begin{document}

\markboth{AbuSiam \& A. Ahriche}{The Scotogenic Model with Two Inert Doublets:
Parameters Space and Electroweak Precision Tests}
\catchline{}{}{}{}{}

\title{The Scotogenic Model with Two Inert Doublets:
Parameters Space and Electroweak Precision Tests}

\author{Abdelrahman AbuSiam\orcidA{}}
\address{Department of Applied Physics and Astronomy, University of Sharjah,
P.O. Box 27272 Sharjah, UAE.\\
u23101998@sharjah.ac.ae}

\author{Amine Ahriche\orcidB{}}
\address{Department of Applied Physics and Astronomy, University of Sharjah,
P.O. Box 27272 Sharjah, UAE.\\
ahriche@sharjah.ac.ae}

\maketitle

\begin{abstract}
In this work, we study a scotogenic extension of the Standard Model
featuring two inert scalar doublets and three singlet Majorana fermions,
where neutrino masses are generated radiatively at one loop. The lightest
among the Majorana fermions and neutral scalars can serve as dark
matter candidates. We explore the parameter space, considering theoretical
constraints (perturbativity, unitarity, vacuum stability) and experimental
limits (lepton flavor violation, Higgs measurements, electroweak precision
observables). Our analysis identifies regions where sizable Yukawa
couplings naturally arise due to constructive interference in the
scalar sector. Additionally, we estimate the oblique parameters, finding
that only $\Delta T$ is sensitive to charged mass splittings, while
$\Delta S$ and $\Delta U$ remain small across the viable parameter
space. However, 60\% of the viable parameter space is excluded by the recent CMS measurement of the $W$ boson mass, since the shift $\Delta M_W$ depends on the oblique parameters, particularly $\Delta T$ that is sensitive to scalar mass splittings.
\end{abstract}


\ccode{PACS numbers: 03.65.$-$w, 04.62.+v}

\section{Introduction}

The Standard Model (SM) of particle physics has been an exceptionally
successful framework for describing fundamental particles and their
interactions with remarkable accuracy. Despite its achievements, the
SM remains incomplete, as it fails to address several critical questions.
Among these are the origin of non-zero neutrino masses and the nature
of dark matter (DM), which constitutes a significant fraction of the
universe's energy density. These unresolved issues underscore the
necessity for new physics (NP) beyond the SM.

Several SM extensions have been proposed to address these shortcomings
by introducing new particles, symmetries, or interactions. A particularly
compelling framework that simultaneously explains neutrino masses
and DM is the scotogenic model, originally proposed by E. Ma~\cite{Ma:2006km}.
This model generates neutrino masses radiatively at one loop level
while naturally providing stable DM candidates, achieved through new
scalar and fermionic degrees of freedom (dof's) stabilized by a discrete
$\mathbb{Z}_2$ symmetry. Over the years, many variants of the scotogenic
model have emerged, some include additional scalar singlets stabilized
by global $\mathbb{Z}_{4}/\mathbb{Z}_{2}$ symmetries~\cite{Ahriche:2020pwq},
while others extend the particle content by introducing $n_{\eta}$
inert doublets and $n_{N}$ singlet Majorana fermions~\cite{Escribano:2020iqq}.
More elaborate constructions explore scale-invariant versions~\cite{Ahriche:2016cio,Soualah:2021xbn,Ahriche:2023hho},
scalar singlet extensions~\cite{Beniwal:2020hjc}, or embed seesaw
mechanisms such as scoto-seesaw~\cite{Barreiros:2020gxu,Barreiros:2022aqu}.
Other variations include gauged $U(1)_{L_{\mu}-L_{\tau}}$ symmetries~\cite{Han:2019rk},
flavor dependent models~\cite{Chen:2019okl,Han:2019lux}, radiative
Dirac neutrinos~\cite{Wang:2017mcy}, and composite Higgs implementations~\cite{Cacciapaglia:2020psm}.

In this work, we consider a variant of the scotogenic model where
the SM is extended by three singlet Majorana fermions $N_{i}$ ($i=1,2,3$)
and two inert scalar doublets $\Phi_{1,2}$~\cite{Ahriche:2022bpx}.
All new fields are odd under a discrete $\mathbb{Z}_{2}$ symmetry,
which forbids tree-level neutrino mass terms while ensuring the stability
of the lightest $\mathbb{Z}_{2}$-odd particle as a viable dark matter
candidate. To establish the model's viability, its predictions must
remain consistent with precision experimental data. In particular,
electroweak precision tests (EWPT) provide stringent constraints on
new physics through the oblique parameters $S$, $T$, and $U$~\cite{Peskin:1991sw}.
These parameters capture the effects of new fields on gauge boson
propagators via loop corrections. Our extended scalar sector introduces
mass splittings and mixing between charged and neutral states, which
can significantly influence these oblique parameters.

Here, we will confront the parameter space allowed by different theoretical
and experimental constraints with the requirements from oblique parameter
bounds. This paper is organized as follows: In Section~\ref{sec:model},
we present the model and review the relevant constraints. Section~\ref{sec:oblique}
derives the contributions to the oblique parameters in our framework.
Section~\ref{sec:results} presents the numerical analysis of our
parameter scan. Finally, we conclude in Section~\ref{sec:conclusion}
with a summary of our findings and their implications.

\section{Model: Neutrino Mass \& Constraints~\label{sec:model}}

In the Lagrangian, we introduce three singlet Majorana fermions $N_{i}$
($i=1,2,3$) and two inert scalar doublets $\Phi_{1,2}$. Explicitly,
the relevant parts of the Lagrangian is given by~\cite{Ahriche:2022bpx}
\begin{equation}
\mathcal{L}\supset\overline{L}_{\alpha}.\epsilon.\left(h_{\alpha,k}\,\Phi_{1}+h_{\alpha,k+3}\,\Phi_{2}\right)N_{k}+\frac{1}{2} \overline{N}_{k}^{C}M_{k}N_{k}+\text{h.c.},
\end{equation}
where $\epsilon=i\sigma_{2}$, $h_{\alpha,k}$ represent the Yukawa
couplings linking leptons, scalars, and fermions, and $M_{k}$ denote
the Majorana masses of the singlet fermions.

The neutrino mass is generated radiatively via the Yukawa interactions
through 12 distinct one loop diagrams.

\begin{figure}[h]
\begin{centering}
\includegraphics[width=0.48\textwidth]{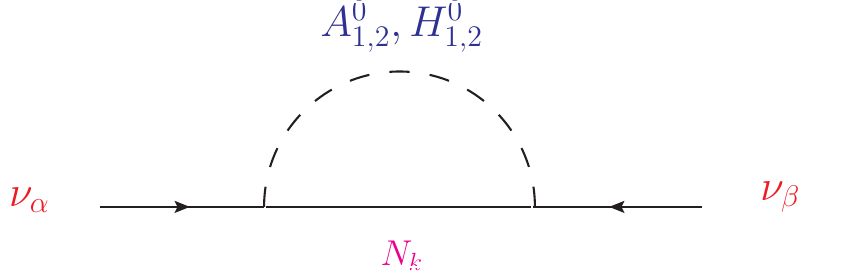} 
\par\end{centering}
\caption{The 12 Feynman diagrams responsible for neutrino mass generation.}
\label{fig:ms-1} 
\end{figure}

The resulting neutrino mass matrix elements are given explicitly by:
\begin{align}
m_{\alpha\beta}^{(\nu)} & =\sum_{k=1}^{3}\sum_{j=1}^{2}M_{k}\left[g_{\alpha k}^{(j)}g_{\beta k}^{(j)}\mathcal{F}\left(\frac{m_{H_{j}^{0}}}{M_{k}}\right)-f_{\alpha k}^{(j)}f_{\beta k}^{(j)}\mathcal{F}\left(\frac{m_{A_{j}^{0}}}{M_{k}}\right)\right],\\
g_{\alpha k}^{(1)} & =c_{H}h_{\alpha,k}+s_{H}h_{\alpha,k+3},\,\,g_{\alpha k}^{(2)}=-s_{H}h_{\alpha,k}+c_{H}h_{\alpha,k+3},\nonumber \\
f_{\alpha k}^{(1)} & =c_{A}h_{\alpha,k}+s_{A}h_{\alpha,k+3},\,\,f_{\alpha k}^{(2)}=-s_{A}h_{\alpha,k}+c_{A}h_{\alpha,k+3},\nonumber 
\end{align}
where the loop function is defined as $\mathcal{F}(x)=\frac{x^{2}\log x}{8\pi^{2}(x^{2}-1)}$.
Here, $H_{1,2}^{0}$ and $A_{1,2}^{0}$ represent the CP-even and
CP-odd eigenstates of the neutral part of the doublets $\Phi_{1,2}$.
The scalar dof's in the doublets $\Phi_{1,2}$ consist of 2 CP-even
scalars $S_{1,2}^{0}\stackrel{\theta_{H}}{\to}H_{1,2}^{0}$, 2 CP-odd
scalars $Q_{1,2}^{0}\stackrel{\theta_{A}}{\to}A_{1,2}^{0}$, and 2
charged scalars $S_{1,2}^{\pm}\stackrel{\theta_{C}}{\to}H_{1,2}^{\pm}$
where the mass eigenstates are obtained via the rotation with the
angles $\theta_{H,A,C}$. The effective couplings, $g_{\alpha k}^{(j)}$
and $f_{\alpha k}^{(j)}$, emerge as combinations of the fundamental
Yukawa couplings $h_{a,k}$ and the scalar mixing angles $\theta_{H}$
and $\theta_{A}$. These angles characterize the mass eigenstates
arising from mixing among the scalar fields.

\vspace{0.5cm}
 \textbf{Theoretical Bounds}: the model parameter space must respect
many theoretical constraints such as perturbativity, vacuum stability,
and unitarity. Perturbativity sets upper limits on couplings, requiring
that the scalar quartic couplings and Yukawa interactions remain within
perturbative ranges; $|\lambda_{i}|,|h_{a,k}|^2\leq 4\pi$. The vacuum
stability requires the scalar potential to be bounded from below,
imposing positivity conditions on the scalar quartic couplings alongside
additional cross-coupling conditions as detailed in~\cite{Kannike:2016fmd}.
In addition, the unitarity further restricts the strength of scalar
self-interactions, ensuring that all scattering amplitudes remain
well behaved at high energies. This requirement typically translates
into the condition: $|\Lambda_{i}|<8\pi$, where $\Lambda_{i}$ are
the eigenvalues of the high energy amplitude matrix that is estimated
for the processes $\phi_{i}\phi_{j}\to\phi_{k}\phi_{l}$. Here, $\phi_{i}$
refers to all scalars including the Goldstone bososns that replace
their corresponding gauge bosons. These constraints are elaborated
in details in~\cite{Ahriche:2022bpx}.

\vspace{0.5cm}
 \textbf{Experimental Limits}: the model parameter space must also
fulfill some experimental constraints such as lepton flavor violating
(LFV) processes, negative collider searches, Higgs and gauge bosons
observables, in addition to some cosmological observations such DM
observables. The collider searches, especially from LEP-II data, place
a lower limit on charged scalar masses, $m_{H^{\pm}}>78,\text{GeV}$~\cite{DELPHI:2003eid};
and the precision measurements of Higgs decay channels, such as $h\to\gamma\gamma$
and $h\to\gamma Z$, must match experimental results to within approximately
$10\%$ of the SM predictions~\cite{ATLAS:2012yve}. The cosmological
relic abundance and direct detection (in case of scalar DM) experiments
constrain the DM candidate in the model. Observations from Planck
set the DM relic density $\Omega h^{2}\approx0.12$, with stringent
limits from direct detection experiments shaping the viable parameter
space further~\cite{Planck:2018vyg}. However, in this setup we have
two possible candidate, fermionic DM ($N_{1}$) or scalar DM (the
lightest among $H_{1}^{0}$ and $A_{1}^{0}$). Here, we will not consider
the constraints from DM in our analysis since it deserves an independent
analysis~\cite{next}. This model is also a subject of constraints
from LFV processes, particularly the decay $\mu\to e\gamma$, tightly
restrict the allowed parameter space. The current experimental bounds
set a strong limit, $B(\mu\to e\gamma)<1.5\times 10^{-13}$~\cite{MEGII:2025gzr}.
These constraints are discussed in details in~\cite{Ahriche:2022bpx}.

\section{The Oblique Parameters~\label{sec:oblique}}

The oblique parameters $S$, $T$, and $U$ originate from radiative
corrections to the propagators of the electroweak (EW) gauge bosons
and play a central role in constraining NP scenarios. In the SM, the
masses of the $W$ and $Z$ bosons are related to the EW gauge couplings
$g$ and $g'$ and the Higgs vacuum expectation value $\upsilon$
through the relations $m_{W}=\frac{1}{2}g\upsilon$ and $m_{Z}=\frac{1}{2}\upsilon\sqrt{g^{2}+g'{}^{2}}$.
The weak mixing angle $\theta_{W}$ is then defined via $\sin^{2}\theta_{W}=1-m_{W}^{2}/m_{Z}^{2}$.
These relations hold at tree level, but are modified once quantum
corrections are considered. In particular, the self energy of the
gauge boson two-point functions ($\Pi_{XY}(q^{2})$ with $X,Y=W,Z,\gamma$)
receive contributions from any new field that is charged under the
EW interactions. The oblique parameters are defined as~\cite{Peskin:1991sw}
\begin{align}
S\; & =\;\frac{4\,s_{W}^{2}\,c_{W}^{2}}{\alpha}\Bigl[\Pi'_{ZZ}(0)-\Pi'_{Z\gamma}(0)\Bigr],\quad T\;=\;\frac{1}{\alpha}\,\frac{\Pi_{WW}(0)-\Pi_{ZZ}(0)}{m_{W}^{2}},\nonumber\\
& U\; =\;\frac{4\,s_{W}^{2}}{\alpha}\Bigl[\Pi'_{WW}(0)-\Pi'_{ZZ}(0)\Bigr],
\end{align}
where $s_{W}\equiv\sin\theta_{W}$, $c_{W}\equiv\cos\theta_{W}$,
$\alpha$ is the fine structure constant, and $\Pi'(0)\equiv d\Pi/dq^{2}\bigl|_{q^{2}=0}$.

These combinations of gauge bosons self energy capture the leading
universal effects of NP that enter through loop corrections to the
gauge sector. The parameter $S$ encodes NP effects in the difference
between $Z$ and photon self energy, $T$ measures isospin breaking
effects and custodial symmetry violation by comparing $W$ and $Z$
self energy at zero momentum. The parameter $U$ accounts for additional
momentum dependent contributions, though it is numerically suppressed
and often taken to be zero in phenomenological analyses. These parameters
are highly constrained by global fits to EWPT, making them a stringent
test of any extension to the SM.

In SM extensions featuring additional scalar states that mix with
the Higgs after electroweak symmetry breaking, new contributions to
the $W$ and $Z$ boson self-energies emerge through their couplings
to the gauge bosons. These effects consequently modify the oblique
parameters. Our framework does not involve neutral and/or charged
singlets, but two inert doublets $\Phi_{1,2}$. The components of
these doublets mix among themselves but remain unmixed with the components
of the SM Higgs doublet.

The charged sector includes six dof's $\phi_{i}^{+}=\sum_{a=1,3}U_{ia}\,S_{a}^{+}$,
where $\phi_{i}^{+}$ represents the charged components of SM doublet
(charged Goldstone) and the inert doublets respectively, and $S_{a}^{+}$
are the mass eigenstates. Then, the matrix $U$ is given by 
\begin{equation}
U=\begin{pmatrix}1 & 0 & 0\\
0 & c_{C} & -s_{C}\\
0 & s_{C} & c_{C}
\end{pmatrix},\label{eq:U}
\end{equation}
with $c_{C}=\cos\theta_{C},\,s_{C}=\sin\theta_{C}$ are mentioned above. In the neutral
sector, we have 6 dof's, i.e., three neutral complex component $\phi_{j}^{0}$,
all of them are doublets members $\phi_{j}^{0}=\sum_{b=1,6}V_{jb}\,S_{b}^{0}$,
where $S_{b}^{0}$ are the mass eigenstates. Then, the matrix $V$
is given by\footnote{In both rotations $\phi_{i}^{+}=\sum_{a=1,3}U_{ia}\,S_{a}^{+}$ and
$\phi_{j}^{0}=\sum_{b=1,6}V_{jb}\,S_{b}^{0}$, the first eigenstate
must correspond to the Goldstone mode.} 
\begin{equation}
V=\left(\begin{array}{cccccc}
i & 1 & 0 & 0 & 0 & 0\\
0 & 0 & c_{H} & -s_{H} & ic_{A} & -is_{A}\\
0 & 0 & s_{H} & c_{H} & is_{A} & ic_{A}
\end{array}\right).\label{eq:V}
\end{equation}

These matrices $U$ and $V$ govern the one loop contributions to
the EW gauge bosons self energy. The resulting shifts in the electroweak
precision observables (EWPOs) are estimated in a general SM extension
by doublets, charged and neutral singlets in~\cite{Peskin:1991sw,Grimus:2008nb}.
The oblique parameters $\Delta T$, $\Delta S$, and $\Delta U$ are
given by 
\begin{align*}
\Delta T & =\frac{1}{16\pi s_{W}^{2}m_{W}^{2}}\left\{ \sum_{a=2}^{n}\sum_{b=2}^{m}\bigl|(U^{\dagger}V)_{ab}\bigr|^{2}F(m_{S_{a}^{+}}^{2},m_{S_{b}^{0}}^{2})\right.\\
 & \left.-\sum_{2\le b<b'\le m}\bigl|\Im\,(V^{\dagger}V)_{bb'}\bigr|^{2}F(m_{S_{b}^{0}}^{2},m_{S_{b'}^{0}}^{2})\right\} ,
\end{align*}
\begin{align}
\Delta S & =\frac{1}{24\pi}\Biggl\{(2s_{W}^{2}-1)^{2}\sum_{a=2}^{n}G(m_{S_{a}^{+}}^{2},m_{S_{a}^{+}}^{2},m_{Z}^{2})\quad+\sum_{2\le b<b'\le m}\Im\,(V^{\dagger}V)_{bb'}^{2}\times\nonumber \\
 & \quad G(m_{S_{b}^{0}}^{2},m_{S_{b'}^{0}}^{2},m_{Z}^{2})-2\sum_{a=2}^{n}\log m_{S_{a}^{+}}^{2}+\sum_{b=2}^{m}\log m_{S_{b}^{0}}^{2}\Biggr\},\nonumber \\
\Delta U & =\frac{1}{24\pi}\Biggl\{\sum_{a=2}^{n}\sum_{b=2}^{m}\bigl|(U^{\dagger}V)_{ab}\bigr|^{2}G(m_{S_{a}^{+}}^{2},m_{S_{b}^{0}}^{2},m_{W}^{2})\quad-\ 2(2s_{W}^{2}-1)\times\nonumber \\
 & \sum_{a=2}^{n}G(m_{S_{a}^{+}}^{2},m_{S_{a}^{+}}^{2},m_{Z}^{2})+\sum_{2\le b<b'\le m}\Im\,(V^{\dagger}V)_{bb'}^{2}G(m_{S_{b}^{0}}^{2},m_{S_{b'}^{0}}^{2},m_{Z}^{2})\Biggr\},\label{eq:STU}
\end{align}
where $F(x,y)$ and $G(x,y,Q^{2})$ are standard two-point functions~\cite{Grimus:2008nb}.

By considering the mixing matrices (\ref{eq:U}) and (\ref{eq:V}),
the oblique parameters formulas (\ref{eq:STU}) are written in our
model as 
\begin{align}
\Delta T & =\frac{1}{16\pi s_{W}^{2}m_{W}^{2}}\Big\{ c_{C-H}^{2}F(m_{H_{1}^{+}}^{2},m_{H_{1}^{0}}^{2})+s_{C-H}^{2}F(m_{H_{1}^{+}}^{2},m_{H_{2}^{0}}^{2})+c_{C-A}^{2}F(m_{H_{1}^{+}}^{2},m_{A_{1}^{0}}^{2})\nonumber \\
 & +s_{C-A}^{2}F(m_{H_{1}^{+}}^{2},m_{A_{2}^{0}}^{2})+s_{C-H}^{2}F(m_{H_{2}^{+}}^{2},m_{H_{1}^{0}}^{2})+c_{C-H}^{2}F(m_{H_{2}^{+}}^{2},m_{H_{2}^{0}}^{2})\nonumber \\
 & +s_{C-A}^{2}F(m_{H_{2}^{+}}^{2},m_{A_{1}^{0}}^{2})+c_{C-A}^{2}F(m_{H_{2}^{+}}^{2},m_{A_{2}^{0}}^{2})-c_{A-H}^{2}F(m_{H_{1}^{0}}^{2},m_{A_{1}^{0}}^{2})-s_{A-H}^{2}F(m_{H_{1}^{0}}^{2},m_{A_{2}^{0}}^{2})\nonumber \\
 & -s_{A-H}^{2}F(m_{H_{2}^{0}}^{2},m_{A_{1}^{0}}^{2})-c_{A-H}^{2}F(m_{H_{2}^{0}}^{2},m_{A_{2}^{0}}^{2})\Big\},\label{eq:dt}
\end{align}
\begin{align}
\Delta S & =\frac{1}{24\pi}\Big\{(2s_{W}^{2}-1)^{2}\left[G(m_{H_{1}^{+}}^{2},m_{H_{1}^{+}}^{2},m_{Z}^{2})+G(m_{H_{2}^{+}}^{2},m_{H_{2}^{+}}^{2},m_{Z}^{2})\right]\nonumber \\
 & +c_{A-H}^{2}G(m_{H_{1}^{0}}^{2},m_{A_{1}^{0}}^{2},m_{Z}^{2})+s_{A-H}^{2}G(m_{H_{1}^{0}}^{2},m_{A_{2}^{0}}^{2},m_{Z}^{2})+s_{A-H}^{2}G(m_{H_{2}^{0}}^{2},m_{A_{1}^{0}}^{2},m_{Z}^{2})\nonumber \\
 & +c_{A-H}^{2}G(m_{H_{2}^{0}}^{2},m_{A_{2}^{0}}^{2},m_{Z}^{2})+\log\frac{m_{H_{1}^{0}}^{2}}{m_{H_{1}^{+}}^{2}}+\log\frac{m_{H_{2}^{0}}^{2}}{m_{H_{2}^{+}}^{2}}+\log\frac{m_{A_{1}^{0}}^{2}}{m_{H_{1}^{+}}^{2}}+\log\frac{m_{A_{2}^{0}}^{2}}{m_{H_{2}^{+}}^{2}}\Big\},\label{eq:ds}
\end{align}
\begin{align}
\Delta U & =\frac{1}{24\pi}\Big\{-2(2s_{W}^{2}-1)\left[G(m_{H_{1}^{+}}^{2},m_{H_{1}^{+}}^{2},m_{Z}^{2})+G(m_{H_{2}^{+}}^{2},m_{H_{2}^{+}}^{2},m_{Z}^{2})\right]\nonumber \\
 & +c_{C-H}^{2}G(m_{H_{1}^{+}}^{2},m_{H_{1}^{0}}^{2},m_{W}^{2})+s_{C-H}^{2}G(m_{H_{1}^{+}}^{2},m_{H_{2}^{0}}^{2},m_{W}^{2})+s_{C-A}^{2}G(m_{H_{2}^{+}}^{2},m_{A_{1}^{0}}^{2},m_{W}^{2})\nonumber \\
 & +c_{C-A}^{2}G(m_{H_{2}^{+}}^{2},m_{A_{2}^{0}}^{2},m_{W}^{2})+c_{A-H}^{2}G(m_{H_{1}^{0}}^{2},m_{A_{1}^{0}}^{2},m_{Z}^{2})+s_{A-H}^{2}G(m_{H_{1}^{0}}^{2},m_{A_{2}^{0}}^{2},m_{Z}^{2})\nonumber \\
 & +s_{A-H}^{2}G(m_{H_{2}^{0}}^{2},m_{A_{1}^{0}}^{2},m_{Z}^{2})+c_{A-H}^{2}G(m_{H_{2}^{0}}^{2},m_{A_{2}^{0}}^{2},m_{Z}^{2})+c_{C-H}^{2}G(m_{H_{2}^{+}}^{2},m_{H_{2}^{0}}^{2},m_{W}^{2})\nonumber \\
 & +s_{C-H}^{2}G(m_{H_{2}^{+}}^{2},m_{H_{1}^{0}}^{2},m_{W}^{2})+c_{C-A}^{2}G(m_{H_{1}^{+}}^{2},m_{A_{1}^{0}}^{2},m_{W}^{2})+s_{C-A}^{2}G(m_{H_{1}^{+}}^{2},m_{A_{2}^{0}}^{2},m_{W}^{2})\Big\},\label{eq:du}
\end{align}
with $c_{a-b}=\cos(\theta_{a}-\theta_{b})$ and $s_{a-b}=\sin(\theta_{a}-\theta_{b})$,
with $a,b=H,A,C$.

The formulas (\ref{eq:dt}), (\ref{eq:ds}) and (\ref{eq:du}) represent
the one loop contributions to the EW observables, that stem from mass
splittings and mixing between the charged, CP-even, and CP-odd scalar
states. The loop functions $F$ and $G$ quantify how these mass differences
affect the self energy of the gauge bosons. The scalar mixing angles
enter through combinations such as $\cos(\theta_{a}-\theta_{c})$
and $\sin(\theta_{a}-\theta_{b})$ (with $a,b=H,A,C$), reflecting
how the different scalar sectors are misaligned. These angle differences
directly affect the couplings between scalar mass eigenstates and
gauge bosons, and thus control the size of the contributions to the
oblique parameters. In physical terms, they determine how strongly
each scalar pair contributes to loop corrections. When the mixing
angles are closely aligned, and the deviations in $\Delta T$ and
$\Delta U$ get smaller. Larger misalignments, on the other hand,
typically enhance these effects. As such, the scalar mixing angles
play a crucial role in shaping the model's consistency with the EWPOs.

Three years ago, a precision measurement of the W boson mass by the
CDF Collaboration yielded $M_{W}^{\text{CDF}}=80,433.5\pm9.4~\mathrm{MeV}$~\cite{CDF:2022hxs}.
This result challenged the SM, as it significantly deviated from both
the prediction of the global electroweak fit and the average of other
$M_{W}$ measurements~\cite{LHC-TeVMWWorkingGroup:2023zkn}. The
discrepancy attracted widespread attention, prompting explanations
in numerous SM extensions, such as those in~\cite{Lu:2022bgw,Strumia:2022qkt,Ahriche:2023hho}.
However, recent measurements by CMS~\cite{CMS:2024wbmass} report
a new value of $M_{W}^{\text{CMS}}=80,360.2\pm9.9~\mathrm{MeV}$,
which agrees with SM predictions and sharply contradicts the CDF-II
result. Given this alignment, the CMS measurement can be used to constrain
the oblique parameters, as the shift in the W boson mass depends on
the oblique parameters as~\cite{Peskin:1991sw,Haller:2018nnx} 
\begin{align}
\Delta M_{W}^{2}= & \,\frac{\alpha}{c_{W}^{2}-s_{W}^{2}}\Big(-\frac{1}{2}\Delta S+c_{W}^{2}\Delta T+\frac{c_{W}^{2}-s_{W}^{2}}{4s_{W}^{2}}\Delta U\Big),\label{eq:mw}
\end{align}
where $\alpha$ is the electromagnetic fine structure constant at
the $Z$ pole.

\section{Numerical Results~\label{sec:results}}

In our analysis, we perform a comprehensive numerical scan over the
model free parameters, while taking into account all relevant theoretical
and experimental constraints discussed above and in~\cite{Ahriche:2022bpx}.
Our focus is on regions of parameter space that makes this model distinguishable
with respect the SM or some of its popular extensions such as the
inert doublet model~\cite{Deshpande:1977rw}. Thus, we focus on the
parameter space regions that are important from collider point of
few, i.e., low Majorana and scalar masses; and non negligible new
Yukawa couplings satisfying $\sum_{ij}|h_{\alpha i}|\geq10^{-4}$.
Our scan is carried out within the following ranges 
\begin{equation}
|\omega_{1,2}|<4\pi,\quad M_{1}<300\,\text{GeV},\quad m_{H_{1}^{0}},\,m_{A_{1}^{0}}<500\,\text{GeV},\quad78\,\text{GeV}<m_{H_{1}^{\pm}}<500\,\text{GeV}.
\end{equation}

\begin{figure}[ht]
\centering
\includegraphics[width=0.95\textwidth]{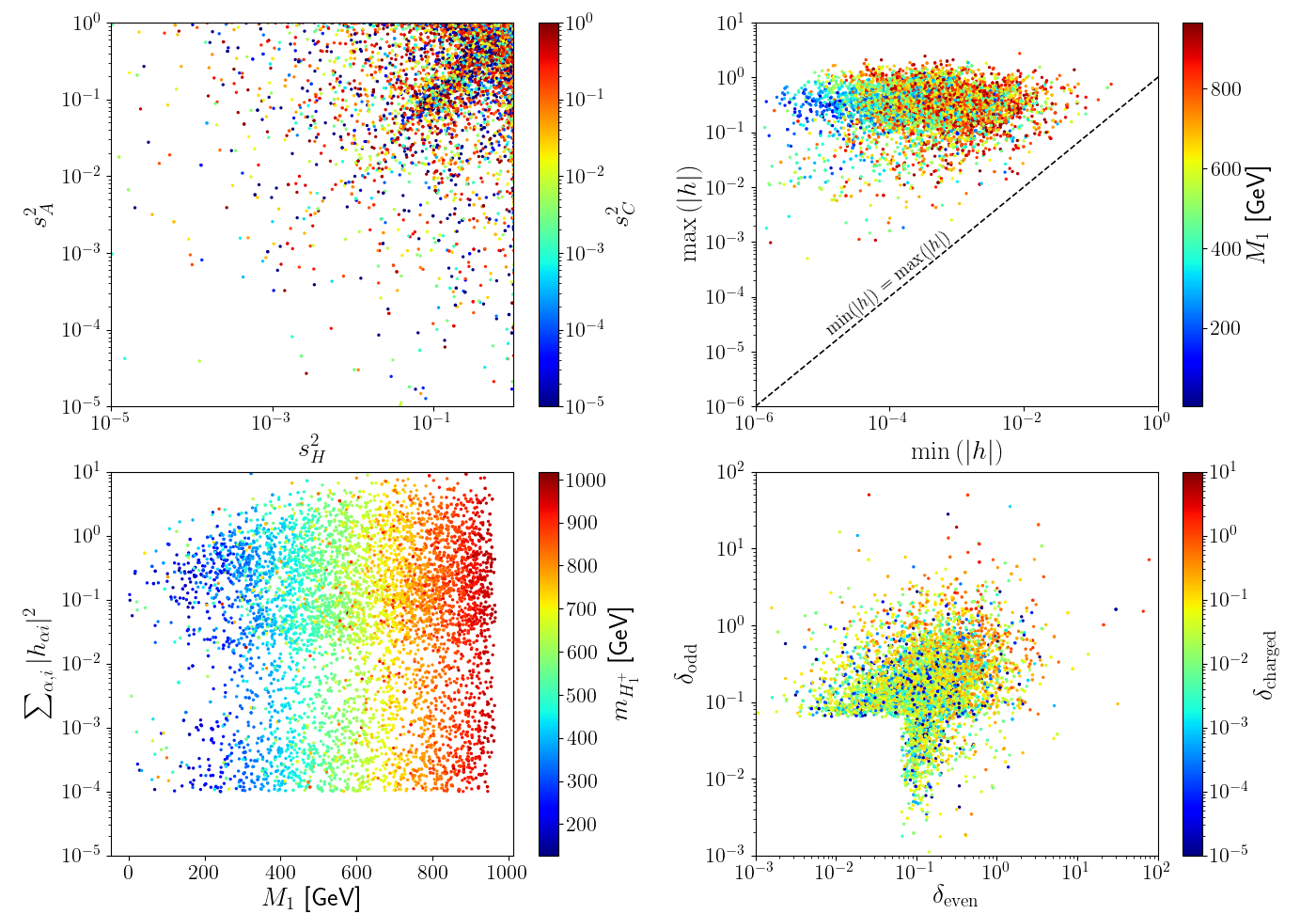}
\caption{\textbf{Top left}: Squared scalar mixing angles $s_{H}^{2}$ vs.\ $s_{A}^{2}$,
colored by $s_{C}^{2}$ for the 5000 BPs, that are satisfying all
theoretical and experimental constraints mentioned above. \textbf{Top
right}: The hierarchy between the minimum and maximum absolute values
of the new Yukawa couplings, $\min(|h_{\alpha i}|)$ vs.\ $\max(|h_{\alpha i}|)$,
where the palette represents the lightest Majorana mass $M_{1}$.
\textbf{Bottom left}: The total Yukawa coupling strength $\sum_{\alpha i}|h_{\alpha i}|^{2}$
vs.\ $M_{1}$, and the mass of the lighter charged scalar $m_{H_{1}^{\pm}}$
shown in the palette. \textbf{Bottom right}: The mass splittings between
scalar states, $\delta_{\text{even}}$ vs.\ $\delta_{\text{odd}}$,
and $\delta_{\text{charged}}$ shown in the palette.}
\label{fig:paramscan} 
\end{figure}

We consider 5000 benchmark points (BPs) that are in agreement with
the constraints mentioned above. These BPs are shown in Fig.~\ref{fig:paramscan}.

Each panel highlights different correlations among the physical parameters
of the model, illustrating how various theoretical and experimental
requirements shape the allowed region. The top-left panel shows the
squared mixing angles $s_{H}^{2}$, $s_{A}^{2}$ and $s_{C}^{2}$,
with most viable BPs clustering near the upper-right region, where
all three scalar mixings are sizable. The top-right panel displays
the hierarchy between the smallest and largest Yukawa couplings, $\min(|h_{\alpha i}|)$
and $\max(|h_{\alpha i}|)$, with the palette indicating the lightest
fermion mass $M_{1}$. The majority of the BPs lie well above the
diagonal line $\min(|h|)=\max(|h|)$, confirming that a strong Yukawa
hierarchy is typical in the allowed parameter space. The bottom-left
panel shows the total strength of the Yukawa couplings $\sum_{\alpha i}|h_{\alpha i}|^{2}$
as a function of $M_{1}$, with the palette corresponds the light
charged scalar mass $m_{H_{1}^{\pm}}$. The allowed BPs span a wide
range in coupling strength, though they are generally more dense around
$\sum|h|^{2}\sim10^{-2}-10^{-1}$, particularly for moderate values
of $M_{1}$. In the bottom-right panel, we show the correlation between
mass splittings $\delta_{\text{even}}=(m_{H_{2}^{0}}^{2}-m_{H_{1}^{0}}^{2})/m_{H_{1}^{0}}^{2}$,
$\delta_{\text{odd}}=(m_{A_{2}^{0}}^{2}-m_{A_{1}^{0}}^{2})/m_{A_{1}^{0}}^{2},$
and $\delta_{\text{charged}}=(m_{H_{2}^{\pm}}^{2}-m_{H_{1}^{\pm}}^{2})/m_{H_{1}^{\pm}}^{2}$.
These splittings may play an important role in determining the EWPOs
and affect significantly the DM relic density as the coannihilation
effect efficient.

\begin{figure}[hbt]
\centering
\includegraphics[width=1\textwidth]{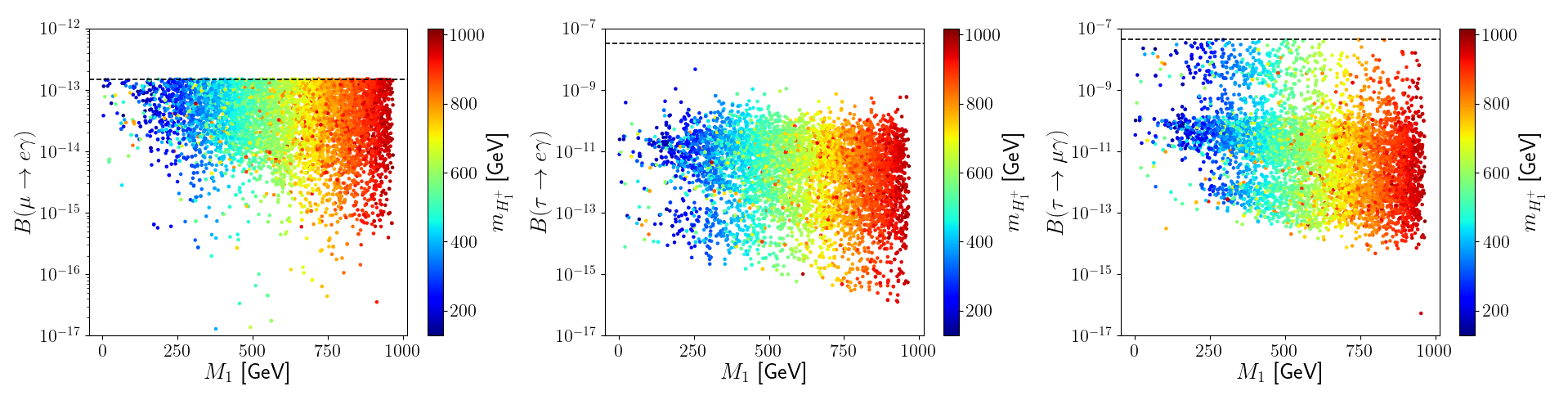}
\caption{The LFV branching ratios for the decays $\mu\to e\gamma$ (right),
$\tau\to e\gamma$ (middle), and $\tau\to\mu\gamma$ (left) as functions
of the lightest fermion mass $M_{1}$, where the light charged scalar
mass $m_{H_{1}^{\pm}}$ is shown in the palette. The horizontal dashed
lines indicate the current experimental upper limits.}
\label{fig:lfv} 
\end{figure}

In Fig.~\ref{fig:lfv}, we present the LFV branching ratios of the
decays $\mu\to e\gamma$, $\tau\to e\gamma$, and $\tau\to\mu\gamma$
as functions of the lightest fermion mass $M_{1}$ and the light charged
scalar mass $m_{H_{1}^{\pm}}$. The definitions of the LFV branching
ratios are given~\cite{Ahriche:2022bpx,Ahriche:2016cio,Chekkal:2017eka}; and the corresponding upper 
bounds are given by $B(\mu \to e\gamma)<1.5 \times 10^{-13}$~\cite{MEGII:2025gzr}, $B(\tau \to e\gamma)<3.3\times 10^{-8}$~\cite{BaBar:2009hkt} and $B(\tau\to\mu\gamma)< 4.2\times 10^{-8}$~\cite{BaBar:2009hkt}.

In Fig.~\ref{fig:lfv}-left, the branching ratio of $\mu\to e\gamma$
is severely fulfilled as a significant region of the parameter space
lies just below the MEG limit of $1.5\times 10^{-13}$~\cite{MEGII:2025gzr}, particularly
for smaller $m_{H_{1}^{\pm}}$ and larger $M_{1}$. However, some
BPs do approach this limit, often corresponding to lighter charged
scalars or larger Yukawa couplings. This indicates the sensitivity
of the process to loop effects from both the scalar and fermion sectors.
In contrast, the middle and right panels, corresponding to $\tau\to e\gamma$
and $\tau\to\mu\gamma$, show that the LFV branching ratios for tau
decays are comfortably below the current experimental bounds of approximately
$10^{-8}$. The results span several orders of magnitude, typically
lying in the range $10^{-13}$ to $10^{-10}$, and exhibit no significant
tension with experimental data.

The color variation in all palettes confirms that larger charged scalar
masses tend to suppress the LFV branching ratios, consistent with
the expected loop suppression behavior.

Overall, The model comfortably satisfies all LFV bounds over most
of the scanned parameter space, particularly when $m_{H_{1}^{\pm}}\gtrsim 200\,\text{GeV}$.
The strongest constraint arises from $\mu\to e\gamma$, which disfavors
scenarios with both light $m_{H_{1}^{\pm}}$ and large Yukawa couplings.

In our model, the Yukawa couplings are given by~\cite{Ahriche:2022bpx}
\begin{equation}
h_{3\times 6}=(U_{\nu})_{3\times 3}.(D_{\sqrt{m_{\nu}}})_{3\times 3}.(T)_{3\times 6}.(D_{(\Lambda'_{i})^{-1/2}})_{6\times 6}.(Q)_{6\times 6}\,\,\,\,\,\,,\label{eq:h}
\end{equation}
with $U_{\nu},\,T$ and $Q$ are dimensionless matrices of order of
magnitude $O(1)$, $m_{\nu}\sim O(0.05)$ represents the experimental
neutrino mass order of magnitude; and $\Lambda'_{i}\propto M_{i}$
are mass dimension parameters that characterize the loop contribution;
and it depends on the Majorana, neutral scalar masses and the scalar
mixing angles $s_{H,A}$. This means 
\begin{equation}
|h|\simeq2\times 10^{-6}\times \left(\frac{\min(\Lambda'_{i})}{10\,\rm{GeV}}\right)^{-1/2}.
\end{equation}
This shows clearly how do the values of $\Lambda'_{i}$ control the
new Yukawa couplings magnitude.

\begin{figure}[ht]
\centering
\includegraphics[width=1\textwidth]{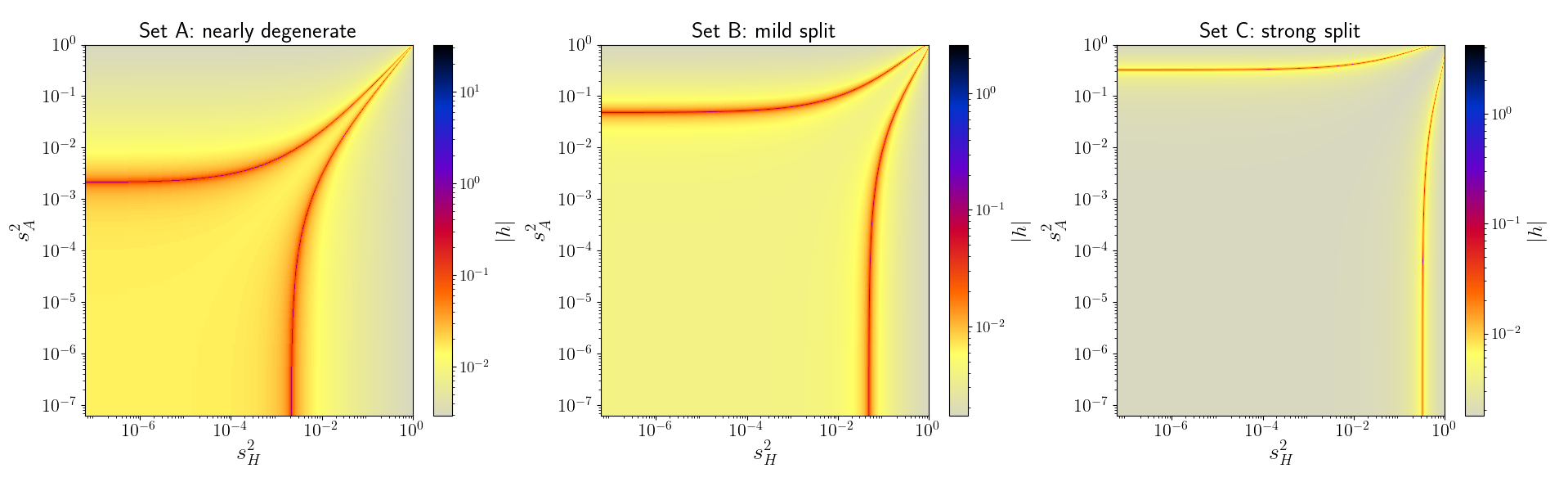}
\caption{The palette shows the Yukawa coupling magnitude $|h|$ across the
scalar mixing plane $(s_{H}^{2},s_{A}^{2})$ for three different scalar
mass spectra, with a color scale indicating the magnitude of $|h|$.
\textbf{Left (set \textbf{A}):} nearly degenerate scalars; \textbf{Middle (Set
B):} mild mass splitting; \textbf{Right (set \textbf{C}):} strong mass splitting.
Darker (more intense) regions correspond to larger Yukawa couplings
needed to reproduce the observed neutrino masses.}
\label{fig:rad_efficiency} 
\end{figure}

In Fig.~\ref{fig:rad_efficiency}, we show the Yukawa coupling magnitude
$|h|$ as a function of the scalar mixing angles $s_{H}^{2}$ and
$s_{A}^{2}$ for three scenarios, each evaluated at a almost degenerate
Majoarana fermions masses $M_{i}=100~\rm{GeV},102~\rm{GeV},105~\rm{GeV}$.
We have chosen three sets of the parameters: 
\begin{align}
\text{Set \textbf{A}: } & m_{H_{1}^{0}}=110~\rm{GeV},\quad m_{A_{1}^{0}}=111~\rm{GeV},\quad m_{H_{2}^{0}}=143~\rm{GeV},\quad m_{A_{2}^{0}}=145.4~\rm{GeV},\nonumber \\
\text{Set \textbf{B}: } & m_{H_{1}^{0}}=150~\rm{GeV},\quad m_{A_{1}^{0}}=170~\rm{GeV},\quad m_{H_{2}^{0}}=300~\rm{GeV},\quad m_{A_{2}^{0}}=374~\rm{GeV},\nonumber \\
\text{Set \textbf{C}: } & m_{H_{1}^{0}}=110~\rm{GeV},\quad m_{A_{1}^{0}}=200~\rm{GeV},\quad m_{H_{2}^{0}}=330~\rm{GeV},\quad m_{A_{2}^{0}}=800~\rm{GeV}.
\end{align}

In set \textbf{A}, where the neutral scalars are nearly degenerate, the ($s_{H}^{2},s_{A}^{2}$) plane exhibits sharp resonance ridges. Here, $|h|$ peaks due to the smallness of $\Lambda_{i}'$ (inversely proportional to the eigenvalues), which amplifies the Yukawa couplings. The near-degenerate masses enable coherent loop contributions, maximizing constructive interference. Off-resonance, destructive interference suppresses effects, requiring smaller couplings for the same observables.

In set \textbf{B}, with moderate mass splittings, the resonance ridges broaden, allowing sizable $|h|$ over a wider mixing range. Constructive interference persists, but the less constrained hierarchy means enhancements occur without fine-tuning, and suppression away from resonances is smoother than in set \textbf{A}.

In set \textbf{C}, the large mass splittings decouple heavy states, pushing resonances to extreme mixing angles ($\min(s_{H}^{2},s_{A}^{2}) \gtrsim 0.3$). Maximal $|h|$ requires fine-tuned mixing where light scalars dominate; elsewhere, destructive interference suppresses contributions. The hierarchy inherently limits
constructive effects, narrowing the viable parameter space.

These results demonstrate that large Yukawa couplings naturally arise along distinct resonance structures, whose shapes and locations depend sensitively on both the scalar mass hierarchy and the mixing parameters. The analysis highlights regions of the parameter space where sizable Yukawa couplings can be achieved while remaining consistent with the neutrino mass generation mechanism.

\begin{figure}[ht]
\centering
\includegraphics[width=0.95\textwidth]{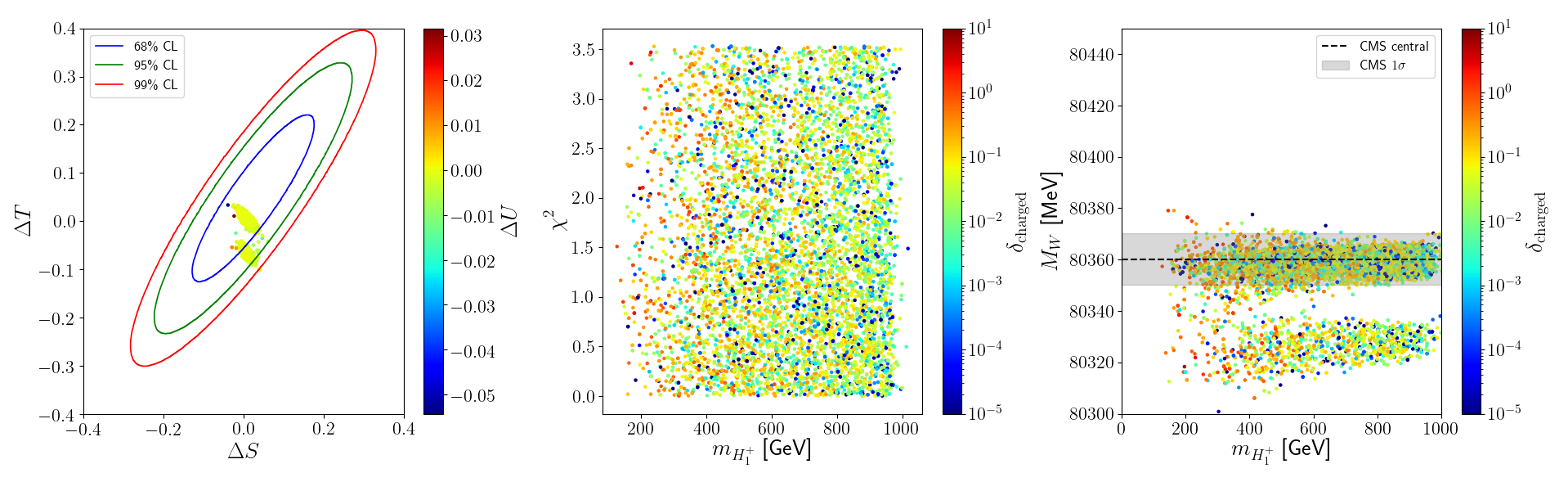}
\caption{\textbf{Left panel:} The oblique parameters $\Delta S$ vs $\Delta T$
vs $\Delta U$, with the 68\%, 95\%, and 99\% confidence ellipses
from global EW fits. \textbf{Middle panel:} EW‐fit $\chi^{2}$ function
vs the light charged scalar mass $m_{H_{1}^{\pm}}$, where the charged
mass splitting $\delta_{charged}$ is shown in the palette. \textbf{Right
panel:} The predicted $W$ boson mass $M_{W}$ as a function of the
light charged scalar mass $m_{H_{1}^{+}}$. The palette represents
the charged mass splitting parameter $\delta_{charged}$. The CMS
central value and its $1\sigma$ uncertainty are indicated by the
dashed line and the shaded band, respectively.}
\label{fig:oblique-scatter} 
\end{figure}

From a phenomenological perspective, the oblique parameters provide
a powerful means to probe mass hierarchies and scalar mixing within
the extended scotogenic model. Fig.~\ref{fig:oblique-scatter} presents
scatter plots illustrating the correlations among the oblique parameters
$\Delta S$, $\Delta T$, and $\Delta U$, the $W$ mass shift and the
$\chi^{2}$ function. The $\chi^{2}$ function is given by~\cite{Arcadi:2019lka} 
\begin{align}
\chi^{2} & =\frac{1}{(1-\rho_{TU}^{2}-\rho_{SU}^{2}-\rho_{ST}^{2}+2\rho_{ST}\rho_{SU}\rho_{TU})}\left(\frac{(\Delta S-\Delta S^{exp})^{2}}{\sigma_{S}^{2}}(1-\rho_{TU}^{2})\right.\nonumber \\
+ & \frac{(\Delta T-\Delta T^{exp})^{2}}{\sigma_{T}^{2}}(1-\rho_{SU}^{2})+2(\rho_{ST}-\rho_{SU}\rho_{TU})\frac{(\Delta S-\Delta S^{exp})(\Delta T-\Delta T^{exp})}{\sigma_{S}\sigma_{T}}\nonumber \\
+ & \frac{(\Delta U-\Delta U^{exp})^{2}}{\sigma_{U}^{2}}(1-\rho_{ST}^{2})+2(\rho_{TU}\rho_{ST}-\rho_{SU})\frac{(\Delta S-\Delta S^{exp})(\Delta U-\Delta U^{exp})}{\sigma_{S}\sigma_{U}}\nonumber \\
+ & \left.2(\rho_{SU}\rho_{ST}-\rho_{TU})\frac{(\Delta T-\Delta T^{exp})(\Delta U-\Delta U^{exp})}{\sigma_{T}\sigma_{U}}\right),
\end{align}
with their experimental values being as follows: $\Delta S=-0.04\pm0.1,\;\Delta T=0.01\pm0.12,\:\Delta U=-0.01\pm0.09$, in addition to the correlation values $\rho_{ST}=0.93, ~\rho_{SU}=-0.70$ and $\rho_{TU}=-0.87$~\cite{ParticleDataGroup:2024cfk}. 
Here, we restrict our scan within $1-\sigma$ level, i.e., $\chi^{2}<3.53$.

In Fig.~\ref{fig:oblique-scatter}-left, we show the oblique parameters
for the 5000 BPs used previously. Most BPs lie within the 95\% confidence
region, indicating that the EW constraints are well respected throughout
the parameter space. The spread is mainly vertical, reflecting the
sensitivity of $\Delta T$ to the mass splittings between charged
and neutral scalars. In contrast, the values of $\Delta S$ are tightly
clustered around zero, since they depend only weakly on the mass splittings,
primarily through logarithmic terms. The palette shows that $\Delta U$
remains significantly small for the BPs majority, that is consistent
with the literature. Fig.~\ref{fig:oblique-scatter}-center illustrates
how does the fit the $\chi^{2}$ function, depending on both the mass
of the lighter charged scalar $m_{H_{1}^{\pm}}$ and the charged mass
splitting. One notices that low $\chi^{2}$ values are achievable
across a broad range of $m_{H_{1}^{\pm}}$ form different charged
mass splittings. For relatively large splittings $\delta_{{\rm charged}}\gtrsim10^{-1}$,
the constraints on $\chi^{2}$ are still fulfilled but only for light
charged scalars $m_{H_{1}^{\pm}}\leq250\,\rm{GeV}$. However,
as the charged scalar mass increases ($m_{H_{1}^{\pm}}\gtrsim300\,\rm{GeV}$),
the fit becomes difficult; and requires small mass splitting $\delta_{{\rm charged}}\lesssim10^{-1}$.
This trend reflects the fact that loop contributions to the oblique
parameters grow with the overall scalar mass scale, making even small
mass splittings more impactful at higher masses. Fig.~\ref{fig:oblique-scatter}-right
shows the correlation between the predicted $W$ boson mass and $m_{H_{1}^{\pm}}$,
with $\delta_{\text{charged}}$ in the palette; compared to the recent
CMS measurement, that is indicated by the dashed line and shaded $1\sigma$
band. One remarks that the CMS measurement is achievable for any mass
range, however, a significant part of the parameter space ($60.6 \%$
of the BPs) is excluded by the CMS $W$ boson mass measurement.

\section{Conclusion~\label{sec:conclusion}}

In this work, we explore the phenomenology of a scotogenic model extended
by two inert scalar doublets and three right-handed Majorana neutrinos.
The model generates neutrino masses radiatively and accommodates sizable
Yukawa couplings along with a rich scalar spectrum. We performed a
comprehensive numerical scan over the parameter space, incorporating
theoretical (perturbativity, unitarity, vacuum stability) and experimental
(EWPT, Higgs decays, LFV processes) constraints. Our scan reveals
that a wide region of the parameter space remains consistent with
LFV bounds. The $\mu\to e\gamma$ decay imposes the strongest constraints,
particularly for light charged scalar masses and large Yukawa couplings.
Nevertheless, most viable scenarios satisfy current experimental bounds.

A key result of our analysis is the identification
of specific regimes in the scalar mixing plane where radiative neutrino
mass generation remains efficient for moderate Majorana and inert
masses, without suppressed Yukawa couplings. When the CP-even and
CP-odd neutral scalars have small mass splittings, the eigenvalues
$\Lambda'_{i}$ that appear in the neutrino mass formula get suppressed
around a sharp resonant structures in the ($s_{H}^{2},s_{A}^{2}$)
plane. Here, large Yukawa couplings $|h|$ emerge due to the $|\Lambda'_{i}|^{-1/2}$
enhancement, rather than purely from constructive interference. As
mass splittings increase, these resonant regions broaden and shift
toward maximal mixing, persisting even in hierarchical spectra. This
behavior highlights how the scalar sector’s structure, governed by
the interplay of masses and mixings, naturally enables viable neutrino
masses across a wide parameter space. Unlike models requiring extreme
fine-tuning or non-perturbative couplings, this mechanism maintains
perturbativity while accommodating sizable Yukawas, reinforcing the
theoretical viability of the scenario.

The scalar sector extension induces non-negligible one loop contributions
to the oblique parameters $S$, $T$, and $U$. Our estimations, incorporating
the full scalar mixing structure, demonstrated that the model can
satisfy the EWPT bounds, with most allowed points lying within the
$95\%$ C.L. region of current global fits. We observe that the $\Delta S$
parameter remains small across the parameter space, while $\Delta T$
is much more sensitive to the mass splittings between charged and
neutral scalars, and that $\Delta U$ remains significantly small
across the parameter space. Interestingly, we found that large charged
mass splitting values are allowed for light charged scalar masses
($m_{H_{1}^{\pm}}\lesssim250$ GeV), while for heavier charged scalars
($m_{H_{1}^{\pm}}\gtrsim300$ GeV), small mass splittings are required
to maintain compatibility with the EWPT. This delicate interplay offers
a powerful indirect probe of the scalar sector at future precision
measurements.

In addition, we have also analyzed the impact of the updated $W$
boson mass measurements. While the earlier CDF-II measurements, that
were in tension with the SM, required large oblique parameters, the
recent CMS measurements are in a good agreement with the SM predictions.
Therefore, this agreement was used here to constraint the oblique
parameters, where we found that about $60\%$ of the viable parameter
space get excluded by the recent $M_{W}$ measurement by CMS. Although
DM constraints lie beyond the scope of this work, the model naturally
supports both fermionic and scalar DM candidates. A comprehensive
study of their relic abundance and direct/indirect detection prospects
remains an important direction for future research~\cite{next}.

\end{document}